\documentstyle[aps,preprint,tighten,prb]{revtex}

\begin{document}

%\draft

\title{PERTURBATIVE EVIDENCE OF NON-UNIVERSALITY IN THE QUANTIZED HALL 
CONDUCTIVITY OF A DISORDERED RELATIVISTIC 2D ELECTRON GAS}
\author{
N. BRALI\'C$^{\dag}$,
C. A. A. DE CARVALHO$^{\ddag}$,
R. M. CAVALCANTI$^{\S,}$\footnote{Present address: 
Institute for Theoretical Physics,
University of California, Santa Barbara, CA 93106-4030, USA.}
and P. DONATIS$^{\ddag}$}
\address{
$^{\dag}$Departamento de F\'\i sica, Pontificia Universidad 
Cat\'olica de Chile \\
Casilla 306, Santiago 22, Chile \\ 
$^{\ddag}$Instituto de F\'\i sica, Universidade Federal do Rio de Janeiro \\
CP 68528, Rio de Janeiro, RJ 21945-970, Brasil \\
$^{\S}$Departamento de F\'\i sica, Pontif\'\i cia Universidade
Cat\'olica do Rio de Janeiro \\
CP 38071, Rio de Janeiro, RJ 22452-970, Brasil}
%\date{\today}
\maketitle

\begin{abstract}

We study a relativistic two-dimensional electron gas in the
presence of a uniform external magnetic field and a random static
scalar potential. 
We compute, in first order perturbation theory,
the averages of the charge density and of the transverse conductivity 
for a white-noise potential, and show that, within this treatment,
their quantized values are modified by the disorder. 
%We discuss the effect of zero vs. finite range correlation length of 
%the random potential.

\end{abstract}

%%%%%%%%%%%%%%%%%%%%%%%%%%%%%%%%%%%%%%%%%%%%%%%%%

\vspace{5mm}

%\section{Introduction}

One of the most puzzling features of the quantum Hall effect (QHE)\cite{QHE} is the
apparent insensitivity of the quantization of the Hall conductivity $\sigma_H$
(in units of $e^2/h$) with respect to type of host material, sample
geometry, presence of impurities or defects, etc.
Being such a high precision phenomenon, it is important to
investigate possible deviations from the quantized values of $\sigma_H$.
So far, theoretical investigations have been concerned mostly with the
effects of disorder,\cite{disorder} and they all seem to agree that the
QHE is robust with respect to it. 

Another possible correction to the QHE is of
relativistic origin: if most of the current is carried by few
electrons because of localization, or if it is carried mainly by electrons
in edge states, there could be relativistic corrections to the QHE
at an accuracy level that could be detected experimentally.
(It should also be noted that the Dirac hamiltonian --- with the velocity of
light replaced with the Fermi velocity --- may be regarded as a low energy 
effective description of an electron-hole system.) 
In this case, too, no corrections to the quantized values of $\sigma_H$
have been found.\cite{MacDonald,Lykken,Zeitlin,DKK}

In this Letter, we study the combined effects of disorder and
relativity. We present a first order perturbative calculation
of the correction to the charge density $j_0$ and the tranverse conductivity
$\sigma_{21}$ due to a random potential with a Gaussian
distribution of zero correlation length. Somewhat surprisingly,
we find that, although the classical relation between the conductivity and the
charge density remains unchanged, {\em their quantized values are modified\/}.
The possible implications of this result are discussed at the end of 
this Letter.

%%%%%%%%%%%%%%%%%%%%%%%%%%%%%%%%%%%%%%%%%%%%%%%%%%%%%%%

%\section{Calculations}

Let us consider an electronic gas in $(2+1)$ dimensions in the presence
of a uniform magnetic field and a random (scalar) potential. Its
Lagrangian is given by ($\hbar=c=1$)
\begin{equation}
{\cal L}=\overline\psi\,(i\partial\!\!\!\slash+eA\!\!\!\slash
-m+\mu\gamma^0)\,\psi.
\label{Lagrangian}
\end{equation}
The field $A=(V,{\bf A})$, where $V$ is a static random
potential, describing quenched disorder, and the
vector potential ${\bf A}$ accounts for the uniform magnetic field
$B$. $\mu$ denotes the chemical potential. The $\gamma$-matrices satisfy
$\{\gamma_{\mu},\gamma_{\nu}\}=2\eta_{\mu\nu}{\bf 1}$, with
$\eta_{\mu\nu}={\rm diag}(1,-1,-1)$ and {\bf 1} 
the $2\times 2$ identity matrix,
and we work with the following representation:
$\gamma_0=\sigma_3$, $\gamma_1=i\sigma_1$, $\gamma_2=i\sigma_2$,
where the $\sigma$'s are the usual Pauli matrices.

The charge density $j_0(x)$ is given by
\begin{equation}
j_0(x) = ie \, {\rm Tr} \, [\gamma_0\,S(x,x)],
\end{equation}
where $S(x,y)$ is the Feynman propagator of the theory, satisfying
\begin{equation}
(i\partial\!\!\!\slash_x+eA\!\!\!\slash(x)-m+\mu\gamma^0)\,S(x,y)=
\delta^3(x-y).
\label{feynman}
\end{equation}
If a perturbing electric field ${\bf E}$ is turned on, a current $J_i(x)$
will be induced which, in the linear response regime, is given by
$J_i(x)=\int d^3y\,\Pi_{i\nu}(x,y)\,{\cal A}^{\nu}(y)$, where 
$\Pi_{\mu\nu}(x,y) = -ie^2\,{\rm Tr}\,[\gamma_{\mu}\,S(x,y)\,
\gamma_{\nu}\,S(y,x)]$ 
is the polarization tensor and 
${\cal A}^{\nu}(x) = (-{\bf E}\cdot{\bf x},{\bf 0})$. 
Thus, the D.C. conductivity tensor, defined as 
$\sigma_{ij}(x) = \lim_{E\to 0}\,\partial J^i(x)/\partial E^j$, 
is given by 
\begin{equation}
\sigma_{ij}(x) = \int d^3y \, \Pi_{i0}(x,y)\,y^j.
\end{equation}

The charge density and the transverse conductivity of the system
without disorder ($V=0$) are well known.\cite{Lykken,Zeitlin,DKK} 
They are given by ($m$ and $eB$ are assumed positive)
\begin{equation}
\sigma_{21}|_{V=0}=-\frac{1}{B}\,
j_0|_{V=0}=\frac{e^2}{2\pi}\,\left\{
\frac{1}{2}-\theta(-\mu-m)+\sum_{n=1}^{\infty}
\left[\theta(\mu-\epsilon_n)-\theta(-\mu-\epsilon_n)\right]\right\}.
\label{sigma_{21}(V=0)}
\end{equation}
where $\epsilon_n\equiv\sqrt{m^2+2neB}$ are the relativistic Landau
levels. Aside from an asymmetry, which is characteristic of the
relativistic theory, Eq.~(\ref{sigma_{21}(V=0)}) exhibits an integer
quantization of the transverse conductivity in units of $e^2/h$.

To investigate the effect of disorder on the system, we average
physical observables over all possible (static) configurations of
$V({\bf x})$, with a suitable weight. Here we choose an
uncorrelated Gaussian probability distribution:
\begin{equation}
\label{P}
P[V]=\exp\left\{-\frac{1}{2g}\int d^2x\,V^2({\bf x})\right\}.
\end{equation}
Since the charge density $j_0$ and the conductivity tensor $\sigma_{ij}$ are
highly non-local functionals of $V$, we shall perform the averaging
perturbatively. This is done by expanding the propagator $S$ in a
power series in $V$ and using the fact that $\langle V({\bf x})\rangle = 0$ 
and $\langle V({\bf x})V({\bf y})\rangle = g \,\delta^2({\bf x}-{\bf y})$
for the distribution (\ref{P}) (averages of products of three or more
$V$'s can be obtained using Wick's theorem, but we shall not need them
in what follows). In matrix notation
($\Gamma(x,x') \equiv e\gamma_0 V(x)\,\delta^3(x-x')$):
\begin{equation}
S = S_0\,\sum_{n=0}^{\infty}(-1)^n(\Gamma S_0)^n.
\label{Dyson}
\end{equation}
Here $S_0$ is the unperturbed Feynman propagator,
satisfying Eq.~(\ref{feynman}) with $V=0$.  It is given by
\begin{equation}
\label{S_0}
S_0(x,y)=M(x,y)\int dp^0\,e^{-ip^0(x^0-y^0)}\,\Sigma(p^0,{\bf x}-{\bf y}),
\end{equation}
where $\Sigma \equiv \Sigma_0\gamma^0 + \Sigma_1\gamma^1 +
\Sigma_2\gamma^2 + \Sigma_3{\bf 1}$, with
($\xi \equiv eB\,({\bf x}-{\bf y})^2/2$)
\begin{mathletters}
\label{Sigma}
\begin{eqnarray}
\Sigma_0(p^0,{\bf x}-{\bf y})&=&
\frac{eB}{8\pi^2}\,e^{-\xi/2}\,
\sum_{n=0}^{\infty}
\left[\frac{p^0+\mu+m}{(p^0+\mu)^2-\epsilon_{n+1}^2} +
\frac{p^0+\mu-m}{(p^0+\mu)^2-\epsilon_n^2}\right]
L_n^0(\xi)\,e^{-\alpha n},
\\
\Sigma_1(p^0,{\bf x}-{\bf y})&=&
-\frac{ie^2B^2}{4\pi^2}\,(x^1-y^1)\,e^{-\xi/2}\,
\sum_{n=0}^{\infty}
\frac{L_n^1(\xi)\,e^{-\alpha n}}{(p^0+\mu)^2-\epsilon_{n+1}^2},
\\
\Sigma_2(p^0,{\bf x}-{\bf y})&=&
-\frac{ie^2B^2}{4\pi^2}\,(x^2-y^2)\,e^{-\xi/2}\,
\sum_{n=0}^{\infty}
\frac{L_n^1(\xi)\,e^{-\alpha n}}{(p^0+\mu)^2-\epsilon_{n+1}^2},
\\
\Sigma_3(p^0,{\bf x}-{\bf y})&=&
\frac{eB}{8\pi^2}\,e^{-\xi/2}\,
\sum_{n=0}^{\infty}
\left[\frac{p^0+\mu+m}{(p^0+\mu)^2-\epsilon_{n+1}^2} -
\frac{p^0+\mu-m}{(p^0+\mu)^2-\epsilon_n^2}\right]
L_n^0(\xi)\,e^{-\alpha n},
\end{eqnarray}
\end{mathletters}
where $L_n^a(z)$ $(a=0,1)$ are Laguerre polynomials,\cite{GR} and
$M(x,y)$ is a gauge dependent factor, given by
\begin{equation}
M(x,y)=\exp\left\{ie\int_{y}^{x}A_{\mu}(z)\,dz^{\mu}\right\},
\end{equation}
where the integral is performed along a straight line
connecting $y$ to $x$. The integral over $p^0$ in Eq.~(\ref{S_0})
must be performed along the contour depicted in Fig.~1.\cite{Chodos,IZ} 
(A UV regulator $e^{-\alpha n}$ is explicitly displayed
in Eq.~(\ref{Sigma}); the limit $\alpha\to 0^+$
must be taken at the end of the calculation of physical quantities.)

Now, we turn to the computation of the first order perturbative
correction to the charge density due to disorder. It is given by
\begin{eqnarray}
\label{j1}
\langle j_0(x)\rangle^{(1)}&=&ie^3\int d^3y\,d^3z\,{\rm Tr}\,[\gamma_0\,
S_0(x,y)\,\gamma_0\,S_0(y,z)\,\gamma_0\,S_0(z,x)]\,
\langle V({\bf y})V({\bf z})\rangle \nonumber \\
&=&4\pi^2i e^3g\int d^2y\int dp^0\,{\rm Tr}\,
[\gamma_0\,\Sigma(p^0,{\bf x}-{\bf y})\,\gamma_0\,
\Sigma(p^0,{\bf 0})\,\gamma_0\,\Sigma(p^0,{\bf y}-{\bf x})].
\end{eqnarray}
Evaluating the trace and performing the integral over {\bf y}, one
finds 
\begin{equation}
\label{j1a}
\langle j_0\rangle^{(1)}=
\frac{ie^5gB^2}{8\pi^3}\,\lim_{\alpha\to 0^+}\,
\sum_{\ell=0}^{\infty}\sum_{n=0}^{\infty}e^{-\alpha(\ell+2n)}
\left[I^+_{\ell+1,n+1}+I^-_{\ell,n}+J^+_{\ell+1,n+1}+
J^-_{\ell,n+1}\right],
\end{equation}
where 
\begin{mathletters}
\begin{eqnarray}
I^{\pm}_{\ell,n}&\equiv&\int \frac{(p^0+\mu\pm m)^3\,dp^0}
{[(p^0+\mu)^2-\epsilon_{\ell}^2]\,[(p^0+\mu)^2-\epsilon_n^2]^2},
\\
J^{\pm}_{\ell,n}&\equiv&\int \frac{2neB\,(p^0+\mu\pm m)\,dp^0}
{[(p^0+\mu)^2-\epsilon_{\ell}^2]\,[(p^0+\mu)^2-\epsilon_n^2]^2}.
\end{eqnarray}
\end{mathletters}
Since the integrands of the above integrals go to zero at infinity at least 
as fast as $(p^0)^{-3}$, one can close the contour depicted in Fig.~1 with a
semicircle of infinite radius in the upper half-plane,
and evaluate the integral using residues. The complete evaluation is
very tedious, but, because of the analytic structure of the integrands,
the result can be written as
\begin{equation}
\label{structure}
\langle j_0\rangle^{(1)}=I_{vac}+I_0\,\theta(-\mu-m)+
\sum_{n=1}^{\infty}\left[I_n\,\theta(\mu-\epsilon_n)+
I_{-n}\,\theta(-\mu-\epsilon_n)\right].
\end{equation}
Here we shall evaluate $I_0$ explicitly; the other $I'$s can be
evaluated similarly.  $I_0$ results from the contribution of the
pole in $p^0=-\mu-m$ to the integral in Eq.~(\ref{j1a}). The terms
containing such pole are $I^-_{\ell,n}$ ($\ell=0$ or $n=0$) and
$J^-_{0,n+1}$ ($n=0,1,2,\ldots$); calculating the residues in
$p^0=-\mu-m$ and collecting the results, we obtain
\begin{eqnarray}
\label{j1c}
\langle j_0\rangle^{(1)}_{-\mu-m}&=&
-\frac{e^5gB^2}{4\pi^2}\,\lim_{\alpha\to 0^+}
\sum_{n=1}^{\infty}
\left[-\frac{e^{-\alpha n}}{2neB}-\frac{m^2e^{-\alpha n}}{n^2e^2B^2}
+\frac{m^2e^{-2\alpha n}}{n^2e^2B^2}+\frac{e^{-2\alpha(n-1)}}{2neB}
\right]\theta(\mu+m)
\nonumber \\
&=& -\frac{e^4gB}{8\pi^2}\,\lim_{\alpha\to 0^+}
\left[\ln (1-e^{-\alpha})-e^{2\alpha}\,\ln(1-e^{-2\alpha})+\frac{2m^2}{eB}
\sum_{n=1}^{\infty}\frac{e^{-2\alpha n}-e^{-\alpha n}}{n^2}
\right]\theta(\mu+m).
\end{eqnarray}
(The factor $\theta(\mu+m)$ assures that the pole $-\mu-m$ contribute
to the integrals only if it is inside the contour of integration.)
Taking the limit $\alpha\to 0^+$, the first two terms combine to
yield
\begin{equation}
\langle j_0\rangle^{(1)}_{-\mu-m}=
\frac{e^4gB\,\ln 2}{8\pi^2}\,[1-\theta(-\mu-m)],
\end{equation}
whereas the third term vanishes (one can take the limit inside the
sum because the latter is uniformly convergent for $\alpha\ge 0$).
It follows that
\begin{equation}
I_0 = -\frac{e^4gB\,\ln 2}{8\pi^2}.
\end{equation}
Performing an analogous calculation for the poles in $p^0=-\mu\pm
\epsilon_n$ $(n=1,2,\ldots)$, one finally obtains
\begin{equation}
\label{j_0^1}
\langle j_0\rangle^{(1)}=\frac{e^4gB\,\ln 2}{8\pi^2}\,
\left\{\frac{1}{2}-\theta(-\mu-m)+\sum_{n=1}^{\infty}
\left[\theta(\mu-\epsilon_n)-\theta(-\mu-\epsilon_n)\right]\right\}.
\end{equation}
(The condition that the pole in $-\mu-\epsilon_n$ $(n=0,1,2,\ldots)$
must be negative to contribute to the integrals in Eq.~(\ref{j1a})
gives rise to a factor $\theta(\mu+\epsilon_n)=1-\theta(-\mu-\epsilon_n)$.  
Thus, such a pole contributes also to $I_{vac}$.)

The first order term in the perturbative expansion of
$\langle\sigma_{21}(x)\rangle$ can be obtained with the help of the
following identity,\cite{Zeitlin,Streda} valid provided
the chemical potential is in an energy gap:
\begin{equation}
\label{sigma-j}
\langle\sigma_{21}\rangle =
-\frac{\partial}{\partial B}\,\langle j_0\rangle.
\end{equation}
From Eqs.~(\ref{j_0^1}) and (\ref{sigma-j}) one immediately obtains
\begin{equation}
\langle\sigma_{21}\rangle^{(1)}=-\frac{e^4g\,\ln 2}{8\pi^2}
\left\{\frac{1}{2}-\theta(-\mu-m)+\sum_{n=1}^{\infty}
\left[\theta(\mu-\epsilon_n)-\theta(-\mu-\epsilon_n)\right]\right\}.
\end{equation}
%

%%%%%%%%%%%%%%%%%%%%%%%%%%%%%%%%%%%%%%%%%%%%%%%%%%%%%

%\section{Discussion of the results}

What is surprising in this result is that it does not vanish
in the low energy limit, although a similar calculation
performed with the non-relativistic propagator gives no
correction to the transverse conductivity. Therefore,
contrary to what one would expect, the average over disorder
does not commute with the non-relativistic limit, at least
for the particular probability distribution we have
considered in this paper. 

On the other hand, if one performs
the calculation presented here using  
\begin{equation}
\label{P'}
P[V]=\exp\left\{-\frac{1}{2gM^2}\int d^2x\,
\left[(\nabla V)^2 + M^2V^2\right]\right\}
\end{equation}
as the probability distribution for the random potential,
one obtains no correction to the transverse conductivity,
regardless of the value of $M$.\cite{Nino} In particular,
one does not recover the results obtained using (\ref{P}) 
if one first averages with (\ref{P'}) and then takes
the limit $M\to\infty$, even though the former distribution corresponds
to the $M\to\infty$ limit of the latter. This seems to indicate
that our results are somewhat pathological.
It should be noted, however, that they emerge from a combination
of disorder, relativity {\em and} perturbation theory.
It would be interesting to investigate if they survive a non-perturbative
treatment. 

%In fact, it is well known that it is necessary to sum
%an infinite number of terms in the pertubative expansion in
%order to lift the degeneracy of the Landau levels and broaden
%them into bands. 

%%%%%%%%%%%%%%%%%%%%%%%%%%%%%%%%%%%%%%%%%%%%%%%%%

\acknowledgments

This work had financial support from Proyecto FONDECYT, under Grant
No.\ 1950794, Fundaci\'on Andes, FUJB/UFRJ, CNPq, FAPERJ, UNESCO and
CLAF.  R.M.C.\ and C.A.A.C.\ thank PUC-Chile, N.B. thanks UFRJ, and
P.D.\ thanks UFRJ and CLAF for the kind hospitality 
while this work was in progress.

%%%%%%%%%%%%%%%%%%%%%%%%%%%%%%%%%%%%%%%%%%%%%%%%%%%%%

%%%%%%%%%%%%%%%%%%%%%%%%%%%%%%%%%%%%%%%%%%%%%%%%%%%%%%%%%%%

\newpage

\noindent
\underline{\bf Figure Caption}:

\vspace{5mm}
\noindent
{\bf Figure 1}: Integration contour in the complex $p^0$-plane used
in the definition of the Feynman propagator.


\begin{references}

\bibitem{QHE} For reviews see, for example,
K. von Klitzing, Rev. Mod. Phys. {\bf 58}, 519 (1986);
R. E. Prange and S. M. Girvin (eds.), {\it The Quantum
Hall Effect\/} (Springer, New York, 1987);
M. Jan{\ss}en, O. Viehweger, U. Fastenrath and J. Hajdu,
{\it Introduction to the Theory of the Integer Quantum Hall
Effect} (VCH Verlagsgesellschaft, Weinheim, 1994).

\bibitem{disorder} There is a vast literature on this subject.
A commented list of references on this and other aspects of
the QHE can be found in:
C. T. Van Degrift, M. E. Cage and S. M. Girvin, {\it Am. J. Phys.\/} {\bf 58},
109 (1990). 

\bibitem{MacDonald} A. H. MacDonald, {\it Phys. Rev.\/} {\bf B28}, 2235 (1983).   

%\bibitem{QED3} J. Schwinger, Phys. Rev. {\bf 82}, 664 (1951);
%A. N. Redlich, Phys. Rev. D {\bf 29}, 2366 (1984);
%V. P. Gusynin, V. A. Miransky and I. A. Shovkovy,
%Phys. Rev. Lett. {\bf 73}, 3499 (1994); Phys. Rev. D {\bf 52}, 4718 (1995);
%Y. Hosotani, Phys. Lett. {\bf B319}, 332 (1993); 
%Phys. Rev. D {\bf 51}, 2022 (1995).

\bibitem{Lykken} J. D. Lykken, J. Sonnenschein and N. Weiss,
{\it Int. J. Mod. Phys.\/} {\bf A6}, 1335 and 5155 (1991).

\bibitem{Zeitlin} V. Zeitlin, {\it Phys. Lett.\/} {\bf B352}, 422 (1995);
{\it Mod. Phys. Lett.\/} {\bf A12}, 877 (1997).
  
\bibitem{DKK} D. K. Kim and K.-S. Soh, {\it Phys. Rev.\/} {\bf D55}, 6218 (1997).

\bibitem{GR} I. S. Gradshteyn and I. M. Ryzhik, {\it Table of
Integrals, Series, and Products\/}, 4th ed. (Academic Press, New York, 1965).

\bibitem{Chodos} A. Chodos, K. Everding and D. A. Owen, 
{\it Phys. Rev.\/} {\bf D42}, 2881 (1990).
%P. Elmfors, D. Persson and B-S. Skagerstam, 
%Phys. Rev. Lett. {\bf 71}, 480 (1993).

\bibitem{IZ} C. Itzykson and J.-B. Zuber, {\it Quantum Field
Theory\/} (McGraw-Hill, Singapore, 1980), p. 91.
  
\bibitem{Streda} P. St\v{r}eda, {\it J. Phys.\/} {\bf C15}, L717 (1982).  

\bibitem{Nino} N. Brali\'c, R. M. Cavalcanti, C. A. A. de Carvalho
and P. Donatis, hep-th/9704091.

\end{references}
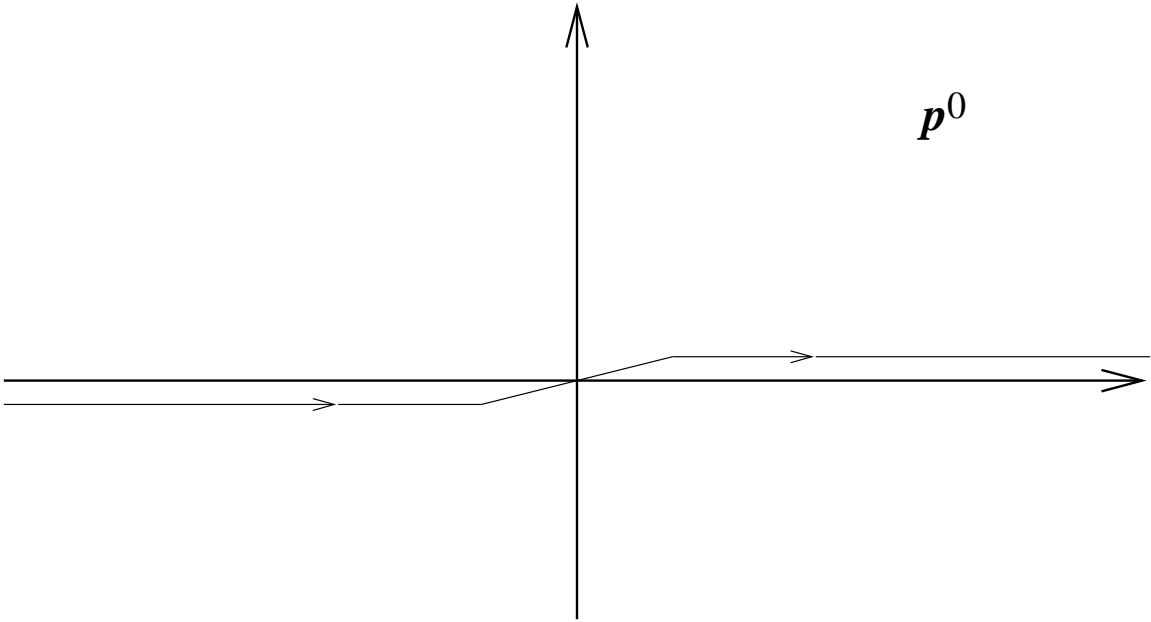
\end{document}